# The Dual Impact of Virtual Reality: Examining the Addictive Potential and Therapeutic Applications of Immersive Media in the Metaverse

Ljubiša Bojić[1,2,3], Ph. D., Senior Research Fellow
(Email address: ljubisa.bojic@ivi.ac.rs; ORCID: 0000-0002-5371-7975)

Jörg Matthes*[,4], Ph. D., Full Professor,
(Corresponding author; Email address: joerg.matthes@univie.ac.at; ORCID: 0000-0002-5371-7975)

Agariadne Dwinggo Samala[5], Ph. D., Assistant Professor
(Email address: agariadne@ft.unp.ac.id; ORCID: 0000-0002-4425-0605)

Milan Čabarkapa[6], Ph. D., Assistant Professor
(Email address: mcabarkapa@kg.ac.rs; ORCID: 0000-0002-2094-9649)

[1] Institute for Artificial Intelligence Research and Development of Serbia; Address of correspondence: 1 Fruskogorska, Novi Sad, Serbia;
[2] University of Belgrade, Institute for Philosophy and Social Theory, Digital Society Lab; Address of correspondence: 45 Kraljice Natalije, 11000 Belgrade, Serbia;
[3] Complexity Science Hub; Address of correspondence: Metternichgasse 8, 1030 Vienna, Austria;
[4] University of Vienna, Faculty of Social Sciences, Department of Communication; Address of correspondence: Währinger Straße 29 (R. 7.47), 1090 Vienna, Austria;
[5] Universitas Negeri Padang, Faculty of Engineering, West Sumatera, Indonesia; Address of correspondence: 482X+WQ West Air Tawar, Padang City, West Sumatra, Indonesia;
[6] University of Kragujevac, Faculty of Engineering; Address of correspondence: 6 Sestre Janjic, 34000 Kragujevac, Serbia;

## Abstract

The emergence of the metaverse—envisioned as a hyperreal virtual universe enabling boundless human interaction—has the potential to revolutionize our conception of media. This transformation could alter society as we know it. This paper identifies addictive features of social media, including immersion, interactivity, real-time access, and personalization. These features are examined within the context of virtual reality through a literature review and content analysis, aimed at exploring the potential consequences of metaverse development. From an initial pool of 193,218 documents, a refined selection of N = 44 relevant papers formed the basis of our qualitative analysis. About half of the analyzed papers indicate that these features contribute to VR addiction. Interestingly, the same features that contribute to addictive behaviors can also be harnessed for positive therapeutic interventions of VR, particularly in treating addictions and managing mental health conditions. This duality, observed in the other half of the papers, emphasizes the complex role of VR technologies, suggesting that they can serve as a substitute for other addictions. This phenomenon is placed into the historical context of evolving media technologies that increasingly mimic reality. The complex interplay of factors contributing to addiction necessitates the development of algorithmic solutions that actively curate diverse offerings, rather than promoting a closed loop of like-minded views. Traditional models of addiction should be adapted to address these unique challenges. Finally, the discussion turned to the implications of these findings for a society where the metaverse is widely accepted as a mainstream technology.

## Introduction

The metaverse's potent capability to provide a digital environment where individuals of diverse geo-flagships can interact, socialize, and work towards a shared goal has turned it into an influential model that could restructure the way we perceive and participate in the world in far-reaching and unprecedented ways (Bailenson, 2021). One of the most significant concerns surrounding the metaverse is its potential to exacerbate the negative side effects associated with social media, with addiction being a primary issue (Cordero, 2024; Kumar et al., 2024; Pal & Arpnikanondt, 2024; Pira, 2024; Bojić et al., 2024; Kim & Kim, 2023; Chen, 2023).

With the rapid convergence of pioneering tech concepts such as VR (Virtual Reality) and AR (Augmented Reality), the development of this substantial virtual cosmos accelerates at a breakneck pace (Bojic, 2022; Samala et al., 2023; Samala et al., 2024). It has therefore become increasingly crucial to apprehend and evaluate the probable aftermath of the metaverse on various key realms such as social interaction and connectivity, creativity, potential addiction formation, multiple aspects of individual happiness, and societal polarization. Discerning how these metamorphoses reflect on society universally and understanding their transformative potential is correspondingly important. By intensively studying the social pertinence of the metaverse, we can gain a more profound comprehension of the prospective future of human bonding and joint effort in a progressively digital era.

The evolutionary chronology of human communication, stretching over a massive 300,000 years, encompasses diverse practices like language development, the creation of cave drawings, use of smoke signals, advancement of printing technology, advent of the fax, the telephone, the internet, social web and now, the next mega milestone, the metaverse (Ning et al., 2021).

### *Approaching the metaverse*

The term "Metaverse," originally conceived by Neal Stephenson in his science fiction novel Snow Crash, outlines a collective digital and cybernetic space birthed from the seamless convergence of physical realities and virtual universes (Stephenson, 1992). It symbolizes a sphere where artificial intelligence (AI), augmented reality (AR), virtual reality (VR), 3D technology, including holography (3D), and Internet of things with NFTs (IoT) amalgamate to architect an immersive and interactive digital world reminiscent of the Internet, yet multiple folds more advanced and captivating (Ning et al., 2021).

The conceptualization of the Metaverse adopts a viewpoint steeped in contemporary technological advancements and progressive development. It envisions an arena where the lines between reality and virtuality blur, where the physical and digital realms intertwine to form an intricate digital ecosystem that transcends conventional boundaries (Bailenson, 2021; Bojic et al., 2024; Samala et al., 2023).

The advent of the Metaverse brings with it a paradigm shift in numerous arenas of research. Pre-existing scholarship largely centers around singular technologies like AR and VR examining their potentials in individual clusters (Han et al., 2022). In contrast to traditional digital spaces or cyberspace, primarily navigated via conventional two-dimensional screens, the Metaverse aspires to construct an environment of complete user immersion. It aims to transform passive observers into cogs of an interactive digital machine that responds and reacts in real-time (Ramesh et al., 2022). It harnesses the potential of 3D technology with the offer of an unrivaled panoramic perspective of the digital cosmos, a prospect scarcely envisaged in traditional research.

Within scholarly circles, the essence of the metaverse is being scrutinized and dissected from various angles. This immersive world will be primarily formed from multi-sensory interactions that are facilitated by virtual reality and will pose significant challenges for designers and architects to draft environments that address cultural sensitivities as well as psychological implications (Ramesh et al., 2022; Mystakidis, 2022).

The Internet of Things (IoT), another transformative technology, plays a crucial role in building the immersive digital universe of the metaverse, which many believe contributes to its high potential for creating addiction (Bojic, 2022). In the case of IoT and NFTs, objects present in both direct and indirect environments (Nadini et al., 2021) can serve as hooks, drawing users back into immersive experiences, especially when these objects evoke feelings of joy, fun, and usefulness, primarily within virtual reality, relegating the direct environment to a secondary position.

Park and Kim's research (2022) revealed that despite certain emotional hurdles connected with avatar design, there are characteristic elements of virtual reality that seem capable of enhancing user experiences beyond realistic expectations. Within the domain of the metaverse, the MMO – mirror mode ontology of real-world mobility and digital doppelgangers of tangible items will contribute prolifically to accentuating the addictive nature of this virtual arena (Mystakidis, 2022; Bolger, 2022). Saker and Frith (2022) sets a goal for the creators of the virtual identities that strike a harmonious equilibrium between enjoyment and existence in reality, as well as effectively mitigating the potential repercussions of a metaverse where individuals might be hesitant to retreat to their real-world lives.

Immersion extends beyond sensory engagement to include a deep absorption into narrative worlds. In narratology, it is a critical component that allows users to experience alternative realities and engage with stories in profound ways (Ryan, 2015; Murray, 1999). Marie-Laure Ryan (2015) explores how virtual environments can create "narrative as virtual reality," where the boundaries between the reader or user and the narrative world diminish, leading to transformative experiences. This exploration of immersion can facilitate empathy, personal growth, therapeutic applications, education, and creativity (Tavinor, 2010; Reyes, 2018; Ryan, 2014).

### *Addictive Nature of Social Media: A Precursor to the Metaverse*

Recent research findings have concluded that new media forms tend to elicit increased levels of addiction as compared to their older counterparts (Kurniasanti et al., 2019; Hoffmann, 2017; Vanden Abeele & Mohr, 2021). This can be attributed to several critical factors that are inherent to new media, namely their vast accessibility, intricate interactivity, and the instant gratification they offer to the users.

Taking a deeper look into the term addiction in this context, it is defined as the state of compulsive use of a medium or platform, wherein the user's sense of control seems fundamentally impaired or compromised (Greenfield, 2015). The new digital technologies and social networking platforms bolster addictive behaviors due to their profoundly immersive nature and interactive designs (Greenfield, 2015).

The constant and effortless availability of new media platforms on portable devices such as smartphones and tablets significantly enhance compulsive engagement, as users have the freedom to access the content at any time and any place, thus leading to an alarming increase in digital indulgence (Greenfield, 2015). A study conducted by Kuss and Griffiths (2017) established that platforms such as social media, online gaming, and various smartphone applications have been instrumental catalysts in increasing the prevalence of internet addiction. The primary cause underlying this predisposition towards compulsiveness seems to be the rewarding effects of instant feedback (Sun et al., 2015), a perceived sense of strong social connection, and an overarching fear of missing out (FOMO) that these engaging platforms typically provide.

In addition to these findings, an essential piece of research (Andreassen et al., 2016) suggested that the design elements of new media platforms significantly fuel the tendency towards addiction. For instance, platforms that offer continuous and unconstrained access to an array of new content, possess the capability for a high degree of customization and personalization, and provide real-time feedback to users not only create an addictive environment but also augment it over time. The potential of receiving affirmative responses trigger dopamine releases, which further contributes to making newer media platforms more addictive (Turel and Serenko, 2012). This cyclical pattern of seeking and receiving validation through immediate feedback can lead to fluctuations in self-confidence and mood alterations (Turel and Serenko, 2012). This issue is further exacerbated by immersive technologies like virtual reality, augmented reality, and live streaming platforms (Granic et al., 2014; Bridgstock et al., 2018). These platforms pose potent challenges for users attempting to disentangle themselves from these captivating environments and return to the real world (Greenfield, 2015).

Unlike their modern counterparts, traditional media forms, such as television, radio, and print media, cannot offer the same degree of interactivity, ease of accessibility or immediate gratification, making them significantly less addictive. On the other hand, user perception of social connectivity plays a vital role in new media platform usage, particularly in the case of social networking sites (Ellison et al., 2007). These user-friendly platforms cater to the innate human need for social validation and a sense of belonging to a community or a group (Ellison et al., 2007).

The fear of missing out, popularly termed as FOMO[1], intensifies the addictive potential of new media platforms. FOMO can lead to obsessive checking behaviors and increased engagement with these platforms since users fear the possibility of missing rewarding experiences enjoyed by others during their absence (Przybylski et al., 2013). This underlying sense of social anxiety coupled with the need for constant connection feeds the addictive behavior associated with widespread new media usage (Dhir et al., 2018).

[1] Though both terms, Gratification and FOMO (Fear of Missing Out), are related to psychology, they are not the same (Przybylski et al., 2013). Gratification in psychology relates to the idea of gaining pleasure, satisfaction, or a sense of fulfillment from something. It's about a feeling of reward or happiness derived from an action or event. FOMO, on the other hand, is a psychological phenomenon where an individual has anxiety about missing out on something fun, fulfilling or rewarding. It's considered a form of social anxiety, where one fears that others may be having rewarding experiences from which they are absent. The relationship between the two is that the fear of missing out might drive individuals to seek gratification through various means to alleviate this fear. However, the emotions and behaviors associated with each are different.

New media platforms are designed to provide users with ongoing access to novel content and utilize complex algorithms that aim to maximize user engagement (Rieder, 2020). These scientifically calculated algorithms consider user preferences and behavioral patterns to deliver custom-manufactured content, which further enhances the appeal of these platforms (Takac et al., 2018). This incessant inflow of curated content fuels the user's desire for more information and content, often leading to feelings of information overload (Carr, 2011).

This unique platform feature of personalization, coupled with immediate feedback mechanisms play a significant role in contributing to the addictive potential that is often correlated with new media platforms (Turel and Serenko, 2012). Although aimed at assisting users in recommending relevant content, products, or social connections, leveraging user preferences, browsing history, and other qualitative information recommender systems on online social media have been linked to exacerbating social polarization, limiting individual information environments through occurrence of filter bubbles[2] and causing platform addiction (Reiter et al., 2022; Ricci et al., 2010; Cao et al., 2020; Siles & Meléndez Moran, 2021; Zhang et al., 2019).

Based on the aforementioned findings regarding the addictiveness of social media, we have compiled a summary of the features that foster technological addiction, the underlying processes driving these behaviors and technologies used in metaverse that correspond with these processes in Table 1.

Table 1. Features of social media, processes contributing to social media addiction and corresponding technologies in metaverse.

| Feature | Process | Related technology in metaverse |
|---|---|---|
| Immersion | Deep mental absorption into the content and high emotional attachment | Augmented Reality (AR), Virtual Reality (VR), 3D technology including hologram (3D), and Internet of Things (IoT), NFTs |
| Interactivity | Immediate stimulation of the cognitive and affective system, Constant state of expectation to receive content and constant state of readiness to react | |
| Real time access (Gratification and FOMO) | Permanenet social gratification and connection, continuous access to new content, independent of time and space | |

[2] The "filter bubble" phenomenon highlights how algorithms integrated into recommender systems limit users' exposure to contrasting perspectives (Reiter et al., 2022). Highly tailored content delivery reinforces pre-existing beliefs and preferences, fostering "echo chambers" where like-minded interactions and confirmation biases thrive (Cinelli et al., 2021). Factors contributing to this include the systems' tendency to prioritize content that stimulates strong emotional responses, further exposing users to charged content that aligns with their beliefs (Reiter et al., 2022). Research shows that societal network architecture can amplify these effects, forming a "self-reinforcing loop" of individuals consuming similar content and solidifying entrenched viewpoints (Reiter et al., 2022). This focus on personalization reduces exposure to novel stimuli, thus narrowing users' digital horizons and interests (Takac et al., 2018; Zeng et al., 2017; Fleder and Hosanagar, 2009).

| | | |
|---|---|---|
| Personalization | High relevance and emotional significance of the content | Recommender system – AI – algorithmic recommendations |

*Objectives of this study*

While researchers and scientists have extensively delved into understanding the potential psychological, physical, and social impacts of virtual reality, the body of scientific research regarding the metaverse is quite limited and needs to be expanded (Han et al., 2022). The relevance of metaverse as a research topic could be justified by the unintended societal impact of previous similar mainstream technologies, such as social media, the issues that have been widely investigated (Krishnan and Zhou, 2019; Scudder, 1999; Gonçalves, 2018; Hagins, 1996; Bojic, 2024; Ali & Zain-ul-abdin, 2021; Dai, Kim, & Jia, 2022; Chagas, 2023). Biocca and Biocca's (2002), Tao et al., (2021) as well as Tracy amd Ashcraft's (2001) work emphasized the need for harmonizing different knowledge bases and building bridges across various fields to effectively navigate the impact of these technologies to society.

In the Metaverse, the influence of computer algorithms on individuals' behavior, cognition, decisions, and perceptions—which is closely tied to personalization—acquires a new level of intricacy. As opposed to the relatively simplified confines of platforms considered in prior studies, the expansive, intricate nature of the Metaverse amplifies the potential reach and transformation that algorithms can bestow. This breeding ground of interactions adds a novel layer of complexity to the established notions, necessitating a new understanding and approach toward modeling, mitigating and leveraging algorithmic effects within the Metaverse (Park & Kim, 2022).

In relation to the identified addictive features of social media—such as immersion, interactivity, real-time access, driven by gratification and fear of missing out, and personalization—we plan to explore these aspects in the context of virtual reality research in the following review. Given that virtual reality is the closest existing medium to the upcoming metaverse[3], our examination will seek to provide valuable insights. Finally, we aim to discuss how these underlying features, which are of pivotal relevance, will significantly impact societies and individuals within the complex landscape of the developing metaverse.

While the metaverse is often associated with VR headsets and immersive hardware, it should be noted that it can also be accessed through browsers and mobile devices. Platforms like Roblox and Decentraland offer 3D virtual environments that users can engage with without specialized equipment. This accessibility broadens the user base and may increase the potential for addictive behaviors, as more individuals can readily immerse themselves in virtual worlds (Jose Miguel et al., 2024). However, this inquiry will focus solely on the use of the metaverse through VR.

[3] Virtual reality refers to a technology that immerses users in a completely digital environment, typically through the use of headsets and sensory devices. It focuses primarily on providing individual experiences by simulating real-world or imagined scenarios, allowing users to interact with and explore these virtual spaces. (Cheremoshkina & Dupina, 2022). The metaverse, on the other hand, is a broader concept that encompasses a vast, interconnected virtual universe where users can not only experience VR but also interact with a persistent, shared digital world. The metaverse integrates various digital technologies, including VR, augmented reality (AR), and the internet, to create a cohesive virtual ecosystem. In this space, users can socialize, work, play, trade, and create using personalized avatars, in an environment that bridges the gap between the virtual and physical worlds (Bojic et al., 2024). While VR is a component of the metaverse, the metaverse itself is a much more expansive and interconnected digital landscape, enabling continuous and multi-faceted user interactions on a global scale.

Therefore, this review aims to answer the following research question:

RQ: What relationship have previous research inquiries identified between the immersive, interactive, real-time, and personalized features of virtual reality and addiction?

Based on the findings, the potential societal impacts of the noted features within the metaverse are discussed, focusing on how these features might shape behavior and social dynamics in this rapidly evolving digital landscape. From a pragmatic standpoint, this study aims to explain potential ethical and regulatory guidelines concerning sensitive issues of addiction that may arise in response to the metaverse. The importance of these considerations is significantly elevated, given the speed of development and the immersive nature of the metaverse (Bojic, 2022).

## Methodology

Our approach encompasses both a thorough literature review and a content analysis, as proposed by Hsieh & Shannon in 2005. An analysis of the substantial volumes of related literature was carried out, predominantly facilitated by the usage of the Scopus (2024). This index base was utilized due to its extensive repository of scientific research, which encompasses a vast array of subjects and topics, making it an invaluable tool for our research endeavor.

In an attempt to investigate deeper into the potential repercussions of certain elements of the metaverse on addiction, we decided to conduct a four separate yet interlinked searches. Each search pertained to one defining feature of the metaverse, such as; immersion, interactivity, real-time access, either gratification or fear of missing out (FOMO), and personalization.

The reason why gratification and FOMO are taken to represent real time access as a feature of metaverse was the fact that these two issues, that basically represent real time access, are mentioned most often in the research inquiries, while “real time access” as such is not mentioned so frequently.

The noted characteristic features of metaverse were then linked with terms “addiction” and “virtual reality” - the latter being chosen due to virtual reality being the closest existing media to the metaverse concept.

This methodology ensured that all relevant literature that mentions virtual reality in conjunction with addiction and at least one aforementioned metaverse percentage was efficiently identified and incorporated into our research study. The specific search terms used for each of the four Scopus searches are highlighted in Table 2.

Scopus (2024) has consistently proven to be a reputable source for a vast assortment of peer-reviewed articles, making it a go-to platform for scholars pursuing highly specific, specialized scientific literature. With its repository of articles, conference transcripts, and patents, Scopus has proven to be an adequate tool for undertaking a detailed and extensive examination of our selected research subject.

From a total of 193,218 documents located within the Scopus index base that mention “virtual reality”, depicted chronologically on Figure 1 in the article title, abstract, or keywords. An intensive search was conducted on the 8th of May 2024 regarding addiction, metaverse, and one of the four metaverse features. As a result of this process, a total of 47 papers were found across four categories (Table 1). On further examination of these 47 papers, one duplicate paper was found (Riva et al., 2019) in the category of Personalization, and two papers (Meng, 2022; Peovic Vukovic, 2010) were found not to be in alignment with our research inquiry. This effectively narrowed our desired literature down to a pool of 44 relevant papers.

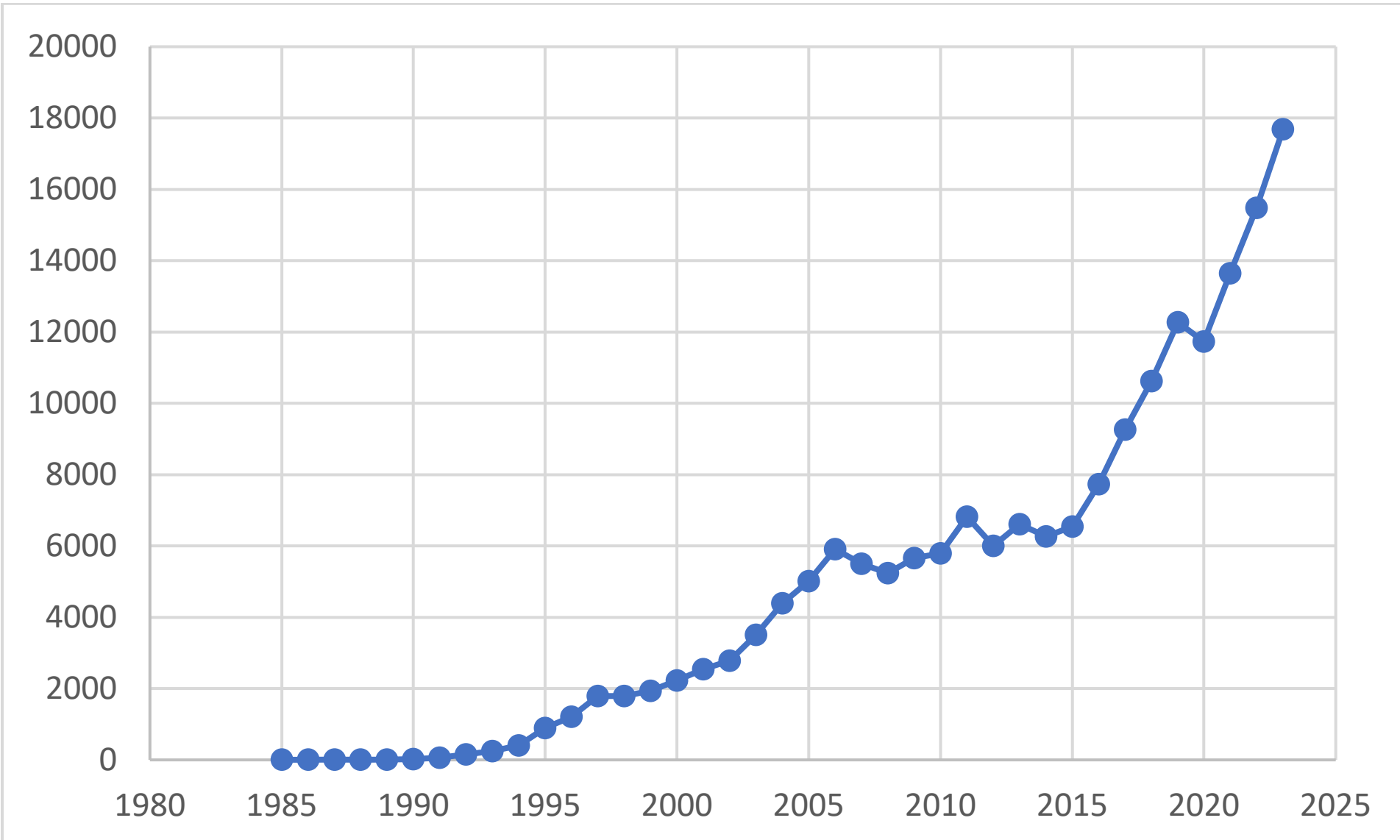

Figure 1. Data derived from the Scopus database from the years spanning 1985 to 2023. The graph portrays an upward trajectory in the utilization of the term "virtual reality" in scholarly inquiries over this period of almost four decades (OSF, 2024).

In terms of the utilization of data within the confines of our research study, we have adopted the use of various resources. These include excel tables containing comprehensive lists of relevant works, word documents containing reference citations in APA formatting, and reference citations in. ris documents. The collected data is freely available via the Open Science Framework (OSF, 2024) to promote the complete transparency of our research methodology. This approach encourages and facilitates additional investigations by academic scholars interested in further exploring this subject matter.

Subsequently, our team carried out a qualitative analysis to appropriately categorize and structure research findings, which are presented in the Results section. For reference purposes, all documents with filenames beginning with “Filtered_” contain references that have been color-coded to indicate their relevance to our study. Those marked in blue indicate they were found to be duplicates or fall outside the thematic scope of our inquiry. Green marked references are specific to addiction to virtual reality, whereas those marked in orange focus on the potential of virtual reality as a treatment strategy for various forms of addiction or pain management.

It has been decided to keep both positive and negative effects of virtual reality to addiction as the large number of analyzed papers actually examines virtual reality as a counter measure or an aiding tool to handle other types of addictions. The other decision made in the process was to preserve studies that analyze virtual reality in the context of gaming and social media, not exclusive to papers that only analyze the use of VR headsets.

Table 2. An overview of the literature review search that was conducted in the Scopus database to retrieve relevant research papers for our study.

| *Metaverse feature* | *Search elements* | *Keywords used in search* | P*apers found though search* | Papers f*ound to be unique and in thematic scope (after filtering)* |
|---|---|---|---|---|
| *Immersion* | Article title, abstract, keywords | Immersion and virtual reality and addiction | 35 | 33 |
| *Interactivity* | Article title, abstract, keywords | Interactivity and virtual reality and addiction | 4 | 4 |
| *Real time access (Gratification and FOMO)* | Article title, abstract, keywords | FOMO or gratification and virtual reality and addiction | 4 | 4 |
| *Personalization* | Article title, abstract, keywords | Personalisation or personalization and virtual reality and addiction | 4 | 3 |
| | | ***Total*** | **47** | **44** |

**Results**

In this section, we investigate into the outcomes of the review and analysis of research papers that were conducted on distinct features of the virtual reality, specifically focusing on immersion, interactivity, real-time access and personalization. The research papers were sorted and organized in a way that highlights the thematic scope papers after a systematic filtering process.

Table 3 provides insights into the number of reviewed papers that have found virtual reality (VR) to engender problems of addiction. However, the table also shows papers in which VR has been found to wield the power to serve as a potential tool that could contribute to addressing various forms of addictions or pain.

Table 3 Research Paper Distribution on VR Features: Insights on Addiction and Pain Relief

| *Metaverse feature* | Papers f*ound to be unique and in thematic scope (after filtering)* | *VR found to be problematic in terms of addiction* | *VR found to be useful in countering different types of addictions or pain* |
|---|---|---|---|
| *Immersion* | 33 | 13 | 20 |
| *Interactivity* | 4 | 3 | 1 |
| *Real time access (Gratification and FOMO)* | 4 | 3 | 1 |
| *Personalization* | 3 | 0 | 3 |
| ***Total*** | **44** | **19** | **25** |

Building upon the patterns drawn from the reviewed studies, we now transition to a deeper examination of these findings. The subsequent sections are bifurcated into two main discussions.

The first facet explores into the concerns regarding the effects of virtual reality, particularly how its features may perpetuate addictive behaviors. This section is highlighted by a number of studies dedicated to exploring mechanisms like VR immersion and the consequential ripple effects that manifest on an individual's cognitive, emotional, and social health.

On the flip side, the second strand explores the potential benefits of the VR experience. Despite the potential drawbacks, a surge of studies indicates that VR can also be a tool for good when judiciously utilized - serving as a therapeutic intervention against different types of addictions and even alleviating physical and psychological pain.

## VR found to be problematic in terms of addiction

This section presents an array of studies to explore the link between VR immersion and addictive behavior, and investigate the ripple effects of this potential addiction on one's cognitive, emotional, and social health.

### *Immersion*

The risk of addiction to virtual reality (VR) and immersive technology has been the focus of several studies. Rózsa et al. (2020) found that immersion in computer gaming is related to maladaptive personality functions and a higher rate of gaming activity, suggesting a potential for addiction. This is echoed by Demetrovics et al.’s (2012) work which identified six dimensions to problematic online gaming, including overuse and immersion. Additionally, a study by Park and Hwang (2009) revealed that psychological concepts like presence and flow are significant factors in online game addiction, again highlighting the link between immersion and addictive behavior.

Another study focusing on MMORPGs such as World of Warcraft found that players drawn to the interpersonal dimensions of the game were more prone to experience negative outcomes associated with their computer use, further emphasizing the potential risk of addiction related to immersive online environments (Snodgrass et al., 2011). In the context of public unrest, immersion in social media was also found significantly associated with addiction (Naranjo-Zolotov et al., 2021).

Research on the use of VR for viewing pornography found it could enhance arousal and pleasure, suggesting the potential for VR to exacerbate compulsive use (Elsey et al., 2019). As fot the gambling behavior, one study suggesting the potential for VR to increase gambling behavior (Detez et al., 2019).

Elevated immersion in virtual reality can also lead to other negative implications beyond addiction. Increased screen time has raised concerns about potential cognitive and emotional impacts on children (Caballero-Julia et al., 2024). Cheremoshkina and Dupina's (2022) study identified a correlation between prolonged internet activity and weakened cognitive abilities. A study by Syakir et al. (2024) noted that addiction to AR face filters could lead to increased online social anxiety.

Dependency on videogaming can also be associated with physical aspects, such as body disconnection. Casale, Musicò, and Schimmenti (2021) conducted a study involving 370 online-game players and showed that low body awareness and high body dissociation predicted the symptoms of Internet Gaming Disorder. This indicates a complex, multi-dimensional nature of online game addiction, which incorporates both physical and psychological elements.

The relationship between motivations to play, in-game behaviors and their impact on daily life was investigated by Billieux et al. (2013). They studied World of Warcraft players and found a relationship between certain motives (like advancement and escapism) and problematic use of the game. However, their longitudinal analyses revealed that high involvement in the game is not necessarily connected to a negative impact on daily living. This highlights that addiction may not always manifest in the same way and demonstrates the need to understand the individual's motivations and behaviors to assess accurately the risk of addiction.

Also, Alontseva and Chechurin (2023) found that pattern of intensive digital engagement among adolescents aged 13 to 18 years, which showed their low level of knowledge about digital rights and ways to handle negative manifestations on the internet. This reinforces that harmful consequences, not just addiction, can be an outcome of extensive immersion in digital environments and emphasizes the need for education to promote a healthier usage of digital technology.

Overall, these studies suggest that VR and immersive technology carry a clear risk of addiction and other negative impacts.

*Interactivity*

In the context of our research focusing on the interactivity feature in virtual reality, four key studies were scrutinized. Each provides valuable insights about the relationship between virtual reality, interactivity, and addiction.

The paper by Soh et al. (2018) closely examines the influence that both parents and peers can have on teenagers' intensity of Internet use and conducts of risky online activities. The study concludes that parental influence can outweigh that of peers if parents actively engage in monitoring and mediating the online activities of their teenagers. This suggests that personal, social, environmental controls outside of the virtual environment may affect the potentially addictive nature of interactive virtual experiences.

Another study by Selivanov, V. V. (2021) takes a unique perspective, examining how immersive didactic VR programs can affect mental states among adolescents. The study confirms that highly immersive VR experiences can lead to increased states of activation, excitement, tone, euphoria, while decreasing states of asthenia. Interestingly, this study indicates that even short-term but repeated immersion in highly interactive VR environments could potentially lead to addiction.

Adding to the discourse, Zhai et al. (2020) construct a structural equation model to examine VR consumption behaviors, specifically VR game addiction (VAD). They concluded that VR realism and consumers' cognitive satisfaction played a significant role as mediators with addictive behavior based on VR-perceived experience of consumers, self-efficacy, and interactivity.

*Real Time Access*

Four studies found in the group of papers, which examine the impact of real-time access, including aspects of gratification and Fear of Missing Out (FOMO), in terms of the potential addictive nature of virtual reality and its remedial uses.

A study by Toronto (2009) illustrates how the virtual world's real-time access, defined by immediate connection and gratification, can promote dissociation where individuals retreat from real-world pain to a subjective state in the virtual world where they can exert control and meet their needs. This work highlights how individual involvement in the virtual world, especially when it disrupts work and significant relationships, can serve as a form of dissociation.

Another central paper by Aboujaoude, et al. (2021) expands the discussion, noting that while the inclusion of "Gaming Disorder" in the International Classification of Diseases has been a critical step, it risks ignoring other dimensions of online psychology. The study brings up two prevailing issues that have arisen with the advent of technology, FOMO and "phubbing." FOMO is induced by social media, causing anxiety over the fear of missing out on what peers might be doing. Phubbing refers to the increasingly common practice of ignoring others in a social setting in favor of one's phone. The authors argue that both these issues can have implications for addiction and psychological distress spread across cultures, suggesting a potential risk of increased addiction due to real-time access enabled by VR technology.

In a study by Chen, Chen, & Ross (2010), the authors discuss Massively Multiplayer Online Game (MMOG) dependency, finding that multimedia realism, one that allows for social interaction, acts as an initial factor influencing other mediating factors leading to MMOG dependency. This study provides additional evidence that real-time access and social interaction in virtual realities can potentially contribute to dependency.

Table 4 A systematic overview of key research on the potential risks of addiction associated with immersive VR experiences.

| *Primary Focus* | *Key Findings* | *Author(s)* |
|---:|---|---|
| *Immersion in Gaming and addiction* | Association between immersion in computer gaming and potential for addiction. | Rózsa et al. (2020), Demetrovics et al. (2012), Park & Hwang (2009) |
| *Interpersonal Dynamics in Games and Social Media* | Relationship between the interpersonal dimensions of games and social media, and the propensity for addiction. | Snodgrass et al. (2011), Naranjo-Zolotov et al. (2021) |
| *VR and Hyperarousal* | VR can enhance arousal, pleasure, and states of euphoria, potentially leading to compulsive use. | Elsey et al. (2019), Selivanov, (2021) |
| *Cognitive and Emotional Impacts of Excessive Screen Time* | Correlation between prolonged internet activity, increased screen time and potential cognitive, emotional impacts. | Caballero-Julia et al. (2024), Cheremoshkina and Dupina (2022), Syakir et al. (2024) |
| *Role of Motivations and Behaviors in Gaming Addiction* | Relationship between certain gaming motives, behaviors and potential problematic use. | Billieux et al. (2013) |
| *Digital Engagement Among Adolescents* | Intensive digital engagement among adolescents can lead to harmful consequences, underscoring the need for education on digital rights. | Alontseva and Chechurin (2023) |
| *Role of Parental and Peer Influence* | Parental influence can outweigh that of peers in determining internet use patterns and potential addiction. | Soh, P. C.-H., et al. (2018) |
| *Immersive Interactivity and Addiction* | High immersion and interactivity in VR can lead to addiction, influenced by VR realism and cognitive satisfaction. | Selivanov, V. V. (2021), Zhai, X., et al. (2020) |
| *Real-time Access and Behavioral Impacts* | Real-time access to VR can disrupt relationships and work, enabling dissociation tendencies and other psychological distresses. | Toronto (2009), Aboujaoude, et al. (2021), Chen, Chen, and Ross (2010) |

The research presented in this section illustrates a correlation between highly immersive VR experiences and increased risk of addiction. The research suggests that the level of immersion, the nature of the activity, and the subjective states of individuals significantly influence the potential for addiction. It is crucial to acknowledge the multidimensionality of this issue, with physical, psychological, and social factors playing pivotal roles.

**VR found to be useful in countering different types of addictions or pain**

With its ability to create immersive environments, VR is being extensively researched for its potential role in various therapeutic applications. The focus of this section is to explore the potential contributions of VR in treating different types of addictions, pain, and mental disorders, as significant part of reviewed papers tackles these issues.

*Immersion*

Immersive aspect of virtual reality has been significantly researched for their potential therapeutic applications in various areas including anxiety, phobias, post-traumatic stress disorders, eating disorders, stroke rehabilitation, and more notably, in pain management and addiction treatment (Garrett et al., 2018).

In the field of addiction therapy, VR has been used to examine and illustrate treatment mechanisms of addictive behaviors such as craving, psychophysiology, affective states, cognition, and brain activity (Mazza et al., 2021). For example, VR-based Approach-Avoidance Task (VR-AAT) was found to be successful in reducing daily smoking and other smoking-related health complications (Eiler et al., 2019, 2020, 2021).

VR alcohol refusal training was explored as a method for engaging patients in dialogues with virtual humans to practice realistic alcohol refusal (Langener et al., 2022, 2023). Similarly, VR was used to help individuals with internet gaming disorder manage gaming-related family conflicts and change their problematic gaming behaviors (Shin et al., 2021).

In terms of pain management, VR has been applied to reduce the severity of mild to medium pain in patients (Bratosin et al., 2019). Advancements in VR further facilitated the development of a home-based pain relief program, leading to substantial immediate and legacy pain relief and a decrease in opioid use (Abd-Elsayed et al., 2021).

As for mental disorders, VR has been used to support the treatment of conditions such as anxiety disorders. For instance, a significant feature of VR in treating anxiety disorders and other psychiatric issues is its potential for controlled sensory stimulation delivery (Maples-Keller et al., 2017).

Appreciating the potential of virtual environments created through VR technology, researchers have worked to understand and harness these environments for therapeutic benefits. Work done by Machulska et al. (2020) has focused on developing a VR-based approach to support those trying to quit smoking. The goal of their innovative method is to reduce the appeal or craving of nicotine, supporting users in their bid to achieve a smoke-free life. Along the same lines, addiction therapy, particularly for alcohol-related disorders, can highly benefit from VR technology (Rick Faust et al., 2021). Tailor-made VR simulations provide an immersive environment where patients can practice real-world coping strategies, enhancing the effectiveness of conventional cognitive behavioral therapy.

Adding more depth to this approach, van Gisbergen et al. (2024) showed how coping mechanisms for real-world cravings could be simulated and practiced using VR-based exposure therapy. Their work went a step further by integrating virtual avatars into the therapy, which interestingly were found to heighten feelings of craving. This suggests that the presence of virtual 'others' could possibly make therapy more challenging, and therefore possibly more effective.

In another application of VR technology, the Game Transfer Phenomena framework proposed by Ortiz De Gortari (2016), investigates altered perceptions induced by virtual immersion. VR's influence on internal cognitions and automatic mental processes has far-reaching implications, from education to therapeutic interventions.

The field of VR-assisted psychotherapy has seen significant advancements by integrating biofeedback techniques into the virtual exposure (Hanshans et al., 2021). This union of immersive technology with biofeedback has shown promise in enhancing traditional therapy methods for addiction-related concerns.

However, despite the promise of VR and IVR, careful research design, and proper pilot studies are necessary to improve the clinical significance of the findings and to prevent potential issues related to excessive virtual identity and addiction (de Clara, 2023). Furthermore, the perceived ecological validity of the virtual environment, or how realistic the users find the virtual experience to be, plays a crucial role in the effectiveness of VR and IVR therapies and treatments (Simon et al., 2020).

*Interactivity*

Research concerning VR as a potential treatment for specific addictions or pain has been conducted. Ehret (2021) propose the use of Virtual Reality in action-centered exposure therapy as an aid for those suffering from Alcohol Use Disorder. This paper highlights an exemplary utilization of VR interactivity as a treatment strategy, suggesting that controlled virtual environments can aid in recovery by reducing the situation dependency of the treatment.

*Real Time Access*

A piece by Scherbaum, et al. (2018), presents a different perspective by examining how individuals having a history of heroin addiction perform in a virtual reality delay discounting task. This study highlights the potential of VR in developing unique strategies that incorporate real-time contexts to evaluate and potentially treat addiction disorders.

*Personalization*

Three articles were examined to evaluate the relationship between personalization in virtual reality and its impact on addiction, focusing particularly on its remedial role for specific cases of addiction or pain.

In a review by Riva, et al. (2019), the researchers identify VR as an effective tool for the treatment and assessment of mental health, particularly in the areas of anxiety disorders, weight and eating disorders, and pain management. The implications of the study highlight the potential of VR's personalized and immersive nature to be harnessed as a therapeutic medium with sustainable effects that extend to real-world scenarios.

Building upon this, Cheng and Ebrahimi (2023) delve into the potential of gamified mental health interventions for mental wellness enhancement and palliation of psychological symptoms. Gamification strategies utilize elements and mechanics that are found in games, such as instant feedback, levels, challenges, and achievements, to motivate behavior change, enhance mental health, and overall well-being. This form of digital intervention, with its personalized and interactive nature, holds significant potential as a therapeutic tool, thanks to the further advancement of VR.

Lastly, an intriguing case is put forth by Siriaraya, et al. (2018) in a study examining the design of ReadySetGoals, a gamified mobile application aimed at supporting therapeutic goal setting within a substance addiction treatment context. This paper's findings underline VR's potential in offering customized and goal-oriented strategies to support addiction treatments.

The growing body of research supports the transformative potential of virtual reality, particularly its inherent personalization and interactivity, in shaping the future of digital therapeutics, which is summed up in Table 5.

Table 5 Summary of the potential application of virtual reality in various therapeutic contexts, from addiction therapy to pain management, as revealed by different authors.

| Primary Focus | Key Findings | Author(s) |
|---|---|---|
| *Use of VR in Mental Disorders* | VR's potential for controlled sensory stimulation delivery in treating anxiety disorders and other psychiatric issues; VR identified as an effective tool for the treatment of mental health conditions such as anxiety disorders | Maples-Keller et al., 2017; Riva, et al. 2019 |
| *Use of VR in Addiction Therapy* | VR used to examine treatment mechanisms of addictive behaviors; Use of VR in action-centered exposure therapy for Alcohol Use Disorder; VR allowing the development of unique strategies to evaluate addiction disorders. | Mazza et al., 2021; Ehret, F., et al., 2021; Scherbaum, et al., 2018 |
| *VR Application in Specific Addictions* | VR-based Approach-Avoidance Task found successful in reducing smoking and smoking-related health complications; Exploration of VR alcohol refusal training to practice realistic alcohol refusal; Use of VR to manage gaming-related family conflicts and change problematic gaming behaviors. | Eiler et al., 2019, 2020, 2021; Langener et al., 2022, 2023; Shin et al., 2021 |
| *Use of VR in Pain Management* | VR used to reduce the severity of mild to medium pain in patients; Development of a home-based VR pain relief program leading to significant pain relief and decrease in opioid use. | Bratosin et al., 2019; Abd-Elsayed et al., 2021 |
| *Integrative and Personalized Approaches in VR Therapy* | Integration of VR and biofeedback techniques; VR gamification strategies for mental wellness enhancement; Gamified mobile application (ReadySetGoals) aimed at supporting therapeutic goal setting in substance addiction treatment context. | Hanshans et al., 2021; Cheng and Ebrahimi 2023; Siriaraya, et al. 2018 |

This section summed up research part of the reviewed papers that identifies VR-assisted therapy in treating different types of addiction, pain, and mental disorders. With its highly interactive and personalized nature, VR promises to offer a uniquely tailored treatment that caters to individuals' specific needs and situations.

**Discussion**: **A Historic Outlook on Reality-Mimicking Features in Media**

The dynamic progression of media technologies has emerged as a potent catalyst that has been instrumental in reshaping arrays of aspects, ranging from the patterns of human communication methods to cultural distinctions and broader societal structures in general (Dominick, 2018). We have witnessed defining shifts and transitions in the mechanism of how information is consumed and disseminated among people with the advent and proliferation of significant media technologies such as newspapers, radio, television, smartphones, and the internet (Hilmes, 2013). Venturing into the aspects that bind these media forms together, one discernible common feature stands out - the persistent endeavor to replicate reality as closely as possible. This attempt at mirroring reality is attained primarily through engaging the user's senses and creating a fully immersive experience (Ramesh et al., 2022). These reality-mimicking characteristics could be instrumental in conceiving a more explicit understanding of the heightened incidence of media addiction (Bojic, 2017). Additionally, it could help in outlining predictions about the impact of the forthcoming metaverse (Mystakidis, 2022). This segment will embark to explore historical evolution of these features in significant media technologies and identify potential implications that hold relevance to the broader discourse around this subject.

Newspapers, one of the earliest variants of mass media, offered content in the written form that primarily catered to the consumption of information. Potently engaging the visual sensory pathways, newspapers channelized information to readers through the medium of written text and associated images (Hilmes, 2013). By heavily incorporating written content to convey aspects ranging from narratives to political perspectives and advertising matter, newspapers made the user's experience immersive predominantly via provoking the imaginative interpretation of the reader. Although newspapers were significantly less effective at sensory-engaging and immersion compared to the forms of media that followed, they still played a pivotal role in establishing the bedrock for future models of mass communication.

The advent of the radio marked a significant expansion in the sphere of sensory engagement by introducing auditory stimulation alongside existing means for the visual consumption of information (Meyrowitz, 1985). The radio, despite the absence of visual elements that characterized newspapers, enhanced immersion by providing real-time auditory content, further cementing the listener's sense of connection with the reported events. Radio broadcasts ushered listeners into imaginative realms of scenarios, characters, and settings, resulting in a more immersive experience than what print media could offer (Olson, 2014). The ability of radio to form intimate bonds with listeners laid down the foundation for subsequent, more reality-mimicking forms of media.

Television staged a revolution in the mass media domain by synergizing visual and auditory experiences into a seamless, consumable platform (Lister et al., 2009). The medium amplified sensory engagement and immersion significantly by offering a mix of real-time and pre-recorded content, giving viewers the dual ability to see and hear the consumed information. Endowed with reality-mimicking capabilities, television eventually superseded radio and newspapers, quickly solidifying its position as the most popular mass media platform (Hilmes, 2013). In doing so, television cleared the path for the development of more sensory-engaging and immersive forms of media, combining visual and auditory stimuli into one package.

The emergence and rise of smartphones have significantly enhanced sensory engagement and immersion by integrating numerous interactive features, such as haptic feedback (Pitts et al., 2012), and providing unhindered access to vast amounts of information and capabilities, creating an immersive experience unmatched by previous media forms. Smartphones empower users to communicate, consume media, and explore an interconnected world from their palms, regardless of place or time (Goggin, 2012). This ease of access and the increasing rates of smartphone ownership have led to a surge in media addiction incidents, affecting psychological functioning and individual well-being in both negative and positive ways (Matthes et al., 2020). As possibly the most all-pervasive form of media is actually combination of a smartphone with internet and variety of interactive apps which involves adequate content and curation of it to individual taste of users. This has drastically transformed how individuals engage with and consume information (Flew, 2014). It unfolds sensory engagement through various formats, including text, images, audio, and video, creating a tailored experience for each user (Lister et al., 2009). The introduction of virtual and augmented realities on online platforms enhances the reality-mimicking features of mass media, contributing to media addiction and an increasing reliance on digital experiences that closely resemble real-life scenarios (Bojic, 2022).

Tracking back through history, the evolution trajectory of mass media platforms reveals a continual emphasis on reality-mimicking features, increasing sensory engagement, and enhanced immersivity. Each innovative iteration in media technology builds upon the groundwork laid by its predecessors (Bojic, 2017). This continuum of advancement has left a resounding impact on issues of media addiction, reliance on digital experiences in the lives of younger generations, among others (Bojic, 2022). Taking this historical perspective into the evaluation of the metaverse, a concept tightly nested within the convergence of technologies that embody multisensory interactions with virtual reality, the potential to further obliterate divisions between actual reality and digital experiences is imminent, which unsurprisingly raises profound concerns (Mystakidis, 2022). As per McLuhan's axiom, the medium itself holds the message (McLuhan, 1964), implying that the progression of media and its inherent structures are demonstrated in Figure 2.

However, it is essential to acknowledge that the evolution of media toward more immersive and reality-mimicking forms also opens up avenues for positive developments. The metaverse, with its capacity to create richly detailed virtual environments, can serve as a platform for immersive storytelling and narrative exploration (Ryan, 2015; Murray, 1999). Engaging with narrative content in virtual spaces allows users to experience 'living' in other possible realities, which can expand their horizons and foster empathy toward diverse experiences (Ryan, 2014). Interactive narratives within the metaverse can facilitate learning, cultural exchange, and ethical reflection by placing users in scenarios that challenge their perspectives (Tavinor, 2011). Thus, the metaverse's reality-mimicking features can be leveraged to provide meaningful and transformative experiences that contribute positively to individual and societal development.

Finally, it is important to note that the problematic aspect of reality-mimicking media development lies not in how new media affects a minority of users who engage with it wisely and productively, but in its impact on the majority, which in turn affects the future of all humans. While such media may be beneficial for some individuals, it can be destructive for larger groups unable to resist the temptations of shallow, addictive, and entertainment-rich content curated by recommender systems (Bojic, 2024; Bojic et al., 2024b).

Figure 2. Old and new media are depicted as pictograms, which are followed by icons to illustrate the senses involved while using them, the potential degree of media addiction (severity) they induce (Bojic, 2017), noted with numbers, and the one- or two-way communication they facilitate (arrows). An icon representing a sense of space is included, highlighting a unique feature of the metaverse.

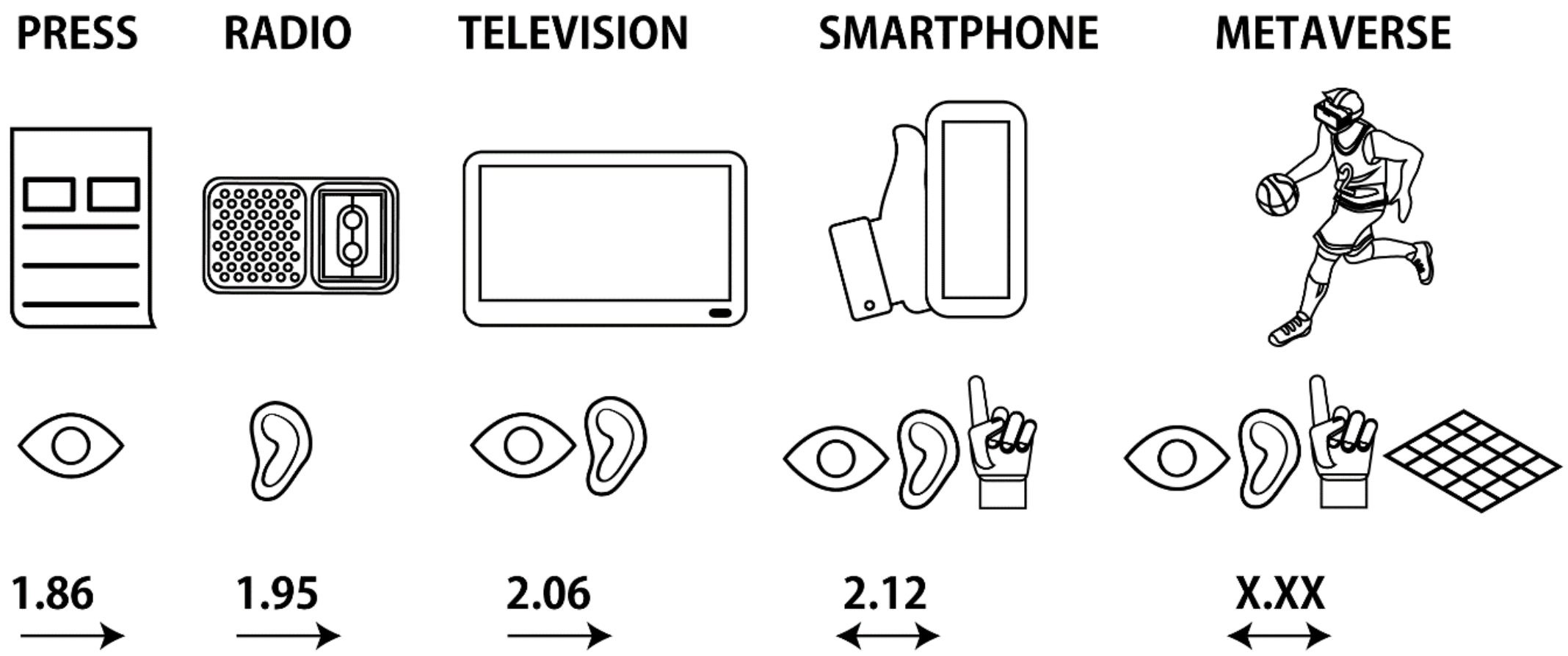

Evaluating modern-day media reveals amplified characteristics that mimic face-to-face communication in the physical sphere than do their ancestral counterparts. This tendency can potentially be associated with contemporary media's inherent addictive quality, as demonstrated by Bojic (2017) study probing into increased levels of media addiction across different types of media. The metaverse is predicted to be the most lifelike, and consequentially, the most addictive medium yet, incorporating elements of spatial awareness along with visual, auditory, tactile, and interactive facets for a truly immersive experience.

## Discussion

The upward trend observed in Figure 1 vividly unpacks the significance and growing prominence of virtual reality in the scientific domain. The number of studies, investigations, and articles referring to virtual reality has increased exponentially. Recognizing this growing interest is crucial as it indicates the significant role that this technological phenomenon could play in shaping our scientific, societal, and cultural landscapes.

In response to our main research question, the review identifies 19 studies that explore how the immersive, interactive, real-time, and personalized features of virtual reality contribute to initiating or increasing addiction specifically to virtual reality itself. Conversely, 25 other studies examine how these same features can be utilized as therapeutic tools to address and potentially alleviate other forms of addiction.

These findings suggest that the metaverse could be so engaging that it might not only be addictive but also serve as a substitute for other addictions. If the new platform proves to be more attractive than reality itself, it could offer an anesthetic experience, introducing a whole new world unimaginable from our current perspective.

Significant societal impacts can be expected if this new iteration of the internet is gradually accepted by the majority, eventually becoming a mainstream tool akin to smartphones today. The metaverse's immersive and interactive features could lead to enhanced frequency and severity of media addiction, making it one of the most addictive media forms to date (Bojic, 2022). Given the exceptional reality-mimicking characteristics of the metaverse, it has the potential to blur the line between the virtual and the real world to an unprecedented degree (Ramesh et al., 2022).

Artificial intelligence, through smartphones and social media platforms, already demonstrates immense power in recommending individually tailored experiences via recommender systems. This allows us to glimpse the potential influence of highly immersive virtual worlds in shaping our entire experience and possibly substituting direct communication.

From a dystopian perspective, humans could end up receiving superficial yet satisfying experiences—a mix of emotions and content tailored to keep them content, though potentially at the expense of the depth and richness of direct communication. While this paper does not delve into whether such a future would truly make people happy or if it is desirable, these are questions worthy of further exploration outside the scope of our current inquiry.

From a more optimistic standpoint, the metaverse's ability to facilitate immersion in alternative realities offers substantial opportunities for enriching human experiences. By enabling users to participate in interactive narratives and virtual worlds, the metaverse can promote creativity, empathy, and learning (Ryan, 2015; Reyes, 2018). As users engage with diverse narratives and perspectives, they can gain insights into different cultures and ways of thinking, contributing to personal growth and social cohesion (Murray, 1999; Ryan, 2014). The therapeutic potential of narrative immersion in virtual environments has been explored in the context of mental health and well-being, suggesting that the metaverse could play a role in healing and self-discovery (Tavinor, 2010). Therefore, while acknowledging the potential risks, it is important to consider the metaverse's capacity to generate positive outcomes by embracing its narrative possibilities. However, it remains crucial to consider how many people will use the metaverse to enhance their personal well-being and how many will do the opposite.

According to the findings, the immersive nature of the metaverse, where users engage in virtual worlds with heightened realism and seamless interaction, would significantly escalate media addiction. This environment, which dissolves the boundaries between virtual and real worlds (Kaplan and Haenlein, 2020), creates a highly addictive space as users become deeply absorbed in virtual experiences and interactions. Several inherent features of the metaverse contribute to this addiction: continuous access to limitless content, real-time social interactions that enhance social belonging, and customizable avatars that allow for self-expression. The traditional distinction between “online” and “offline” may become irrelevant, with interactions in the metaverse being seen as legitimate as those in the physical world. Interaction with AI-generated personalities might become as satisfying and engaging as interactions with real people, further blurring the lines between "real" and virtual beings.

Similar to existing new media platforms, the metaverse has the potential to enrich human interaction, provide novel learning experiences, and improve well-being (Lee, 2022). However, it could also lead to social polarization by creating insulated comfort zones where individuals primarily engage with like-minded users, reinforcing their beliefs and exacerbating divisions. As users become more deeply nested within their virtual communities, they may become insulated from diverse opinions and worldviews. Algorithmic recommenders within the metaverse, which relate to the examined feature of personalization, could perpetuate echo chambers and filter bubbles, steering users toward environments that align with their preferences and shielding them from controversial viewpoints (Reiter et al., 2022).

This groundbreaking platform will impact human relationships, bringing both benefits and detriments. While the metaverse offers opportunities for connection, cooperation, and socialization in captivating virtual environments—fostering profound connections and cross-cultural understanding—it could also strain relationships if users prioritize virtual interactions over face-to-face ones. The metaverse could exacerbate issues like online harassment, cyberbullying, and privacy infringements, posing potential harms to relationships.

The metaverse's immersive capabilities could negatively influence overall happiness and psychological well-being through factors such as addiction, social isolation, and the blurring of virtual and real-world boundaries. The engaging nature of the metaverse may lead individuals to invest significant time and emotions, akin to current addictions to social media and gaming, potentially harming their mental health and creating complex dependencies. While the metaverse promises global integration, it might also create social isolation by reducing opportunities for genuine face-to-face interaction. Users deeply engaged in the metaverse may feel detached from friends and family who prefer real-world experiences, leading to loneliness and negative impacts on mental health.

The hyper-immersive environment of the metaverse can blur the distinction between virtual and real-life experiences, causing confusion and cognitive dissonance. Users may struggle to correlate their virtual achievements and friendships with their real-world existence, leading to dissatisfaction and a decline in life satisfaction. While the metaverse offers unprecedented opportunities, it also presents a mix of potential repercussions that stakeholders must navigate carefully.

*Practical implications*

Our findings indicate a significant potential for addiction within immersive virtual environments like the metaverse. This highlights the urgent need for developers and policymakers to implement safeguards against excessive use. Practical steps could include regulatory changes, app design modifications, and educational initiatives. For example, laws such as the Digital Services Act (Turillazzi et al., 2023) and the EU AI Act (Veale & Zuiderveen, 2021) could introduce requirements to limit usage time and enforce mandatory algorithmic diversity in content recommendations. This could involve a personalized balance between educational and entertainment content, positive and negative sentiments, desired and undesired attitudes, and varying levels of topic interest, to prevent users from becoming isolated in content bubbles and reduce technology addiction (Bojic, 2024; Bojic, 2022). Developers should be encouraged to create features that promote healthy engagement, such as time-tracking tools and notifications about prolonged use. Additionally, governments should implement educational programs to inform users about the risks of addiction associated with immersive technologies.

The research also highlights the dual nature of VR, showcasing its therapeutic potential. Mental health professionals can leverage VR to treat various conditions such as phobias, anxiety disorders, and even addiction to other substances. Integrating VR into therapeutic practices can offer immersive and controlled environments beneficial for various treatment protocols. Thus, fostering research on VR-based therapeutic interventions can bolster mental health care.

Architects and designers of virtual environments should consider the psychological and cultural implications of their designs. Creating spaces that are not only engaging but also promote safe and healthy interactions is vital. This can involve designing environments that encourage physical activity, social interaction, and educational experiences which can help mitigate the potential for addiction.

The role of policymakers and technical professionals in crafting regulations and engineering solutions are imperative to prevent a surge in media addiction and its associated socio-psychological implications. The introduction of regulations that monitor and curb potentially harmful elements within the metaverse, particularly those of a distracting or intrusive nature that may create barriers to users wishing to disengage or withdraw, could also be a crucial step towards establishing a balanced metaverse experience. Also, on the horizon is the potential for the metaverse to contribute to heightened social fragmentation and polarization. This ominous prospect highlights an immediate and vital need for regulation and transformation in the algorithms that operate within the metaverse. It necessitates an approach where we actively curate a diverse offering rather than encouraging a closed loop of like-minded views. For example, Bojic (2024) offers a tripartite decision-making algorithm, equally influenced by individuals, society, and tech corporations. This could necessitate regulatory bodies to play a supervisory role, particularly when it comes to managing potential problems relating to the potential scope and widespread impact of the metaverse.

*What kind of future are we looking into?*

The addictive potential of the metaverse stems from its ability to offer highly immersive experiences that closely mimic or even surpass real-life interactions. Users can become deeply absorbed in virtual environments where they can curate their identities, experiences, and social interactions to an unprecedented degree. This level of immersion can lead to a preference for virtual over physical reality, as the metaverse provides immediate gratification, personalized content, and an escape from the complexities of the real world.

In such a society, there's a risk that individuals may exhibit lower emotional expressivity and intensity. Interactions within the metaverse are mediated through avatars and digital interfaces, which can filter out human emotions conveyed through facial expressions, tone of voice, and body language. Over time, reliance on virtual interactions might erode the ability to perceive and express emotions authentically, leading to superficial connections and a diminished capacity for empathy.

Creativity, too, could be adversely affected. The algorithmic curation of content can confine users within echo chambers of familiar themes and ideas. This personalization may limit exposure to diverse perspectives and novel stimuli that are essential for creative thought. Additionally, the convenience of consuming tailored content might reduce the motivation to create original works, leading to a passive consumption culture rather than active creation.

A society engrossed in the metaverse might also witness a decline in participation in physical reality. Physical activities, outdoor experiences, and face-to-face social interactions could be supplanted by virtual counterparts. This shift has implications not only for individual well-being—affecting physical health and mental resilience—but also for community cohesion and societal structures that rely on real-world engagement.

While virtual reality offers promising therapeutic applications for traditional addictions, this approach raises critical concerns about substituting one dependency with another. The immersive and engaging nature of the metaverse may indeed help individuals overcome addictions to substances like alcohol or drugs. However, it could simultaneously foster a new, potentially stronger addiction to the virtual environment itself. This raises the question of whether such an outcome is desirable. If reliance on the metaverse leads to excessive disengagement from physical reality, diminished emotional expression, and reduced creativity, then merely reducing traditional addictions might not justify encouraging a societal shift toward living predominantly in virtual spaces.

The metaverse's capacity to anesthetize users poses ethical and psychological challenges. The constant availability of pleasurable and controlled virtual experiences might encourage avoidance of real-life challenges and responsibilities. Such escapism can hinder personal growth, reduce resilience, and exacerbate issues like loneliness and depression when individuals are confronted with the less controllable aspects of physical reality.

In order to avoid this future, it is imperative to design virtual environments that encourage meaningful interactions, creativity, and emotional engagement rather than mere consumption.

*Theoretical implications*

This study advances the theoretical understanding of media addiction in the context of highly immersive technologies. It suggests that traditional models of media addiction should be adapted to account for immersion, interactivity, real-time access, and personalization. The blurring line between virtual and real experiences necessitates a redefinition of how we conceptualize user engagement and dependency. The role of personalized content and algorithms in shaping user behavior becomes more pronounced in immersive environments, requiring new theoretical frameworks to understand their impact, as suggested by Kardefelt-Winther (2017).

The findings call for an interdisciplinary approach to studying digital addiction, intertwining insights from psychology, neuroscience, computer science, and social sciences. This will lead to more comprehensive theoretical models that can explain the relationships between media technology design and user behavior (Biocca & Biocca, 2002).

The concept of 'presence' in virtual environments, historically confined to telepresence and VR studies, must be reevaluated in the context of the metaverse. This study suggests that the theory of presence should incorporate the persistent and pervasive nature of engagement in the metaverse, potentially leading to a more dynamic understanding of user experience and attachment (Lombard & Ditton, 2006).

This research adds a significant chapter to the evolutionary narrative of media consumption, showing how each advancement towards more reality-mimicking media (from print to internet to metaverse) progressively increases its addictive potential. This offers a theoretical basis for predicting future trends in media engagement and dependency (Meyrowitz, 1985).

The study's focus on algorithmic curation in the metaverse necessitates a deeper theoretical understanding of how algorithms shape user behavior, decisions, and social dynamics. This extends existing theories on information filtering, echo chambers, and recommendation systems by highlighting their potential effects in fully immersive environments (Pariser, 2012).

The metaverse, by enabling new forms of social interaction and connection, requires a revision of social interaction theories. Theories must now account for how virtual identities, digital avatars, and simulated environments influence real-world social behaviors, relationships, and community formations (Ellison et al., 2007).

The study emphasizes the theoretical significance of gratification and fear of missing out (FOMO) in driving media addiction. It highlights the necessity to further explore these psychological constructs in the context of real-time, immersive digital experiences and their profound impact on user engagement and mental health (Przybylski et al., 2013; Turel & Serenko, 2012).

By demonstrating the cognitive and emotional impacts of prolonged use of the metaverse, the study calls for revised cognitive models that integrate the long-term effects of high-intensity digital interaction. These models should consider how immersive experiences influence attention, memory, learning, and emotional regulation. (Greenfield, 2014; Limone & Toto, 2021).

The findings highlight the need for incorporating ethical theories into technology development. Theories of tech ethics must evolve to address the immersive, pervasive nature of the metaverse and propose guidelines for responsible development and usage (Floridi, 2013).

The dual nature of VR as both a potential source of addiction and a therapeutic tool presents unique theoretical challenges. There is a need to develop theories that explore how the same technologies can have both detrimental and beneficial effects depending on the context and manner of use, thus guiding future research in digital therapeutics. (Maples-Keller et al., 2017).

Future research can be better equipped to understand the complexities introduced by the metaverse and develop more effective interventions, policies, and theoretical models that cater to the evolving digital landscape.

### *Limitations and Future Research*

This review presents several limitations due to its reliance on existing literature. Even though it consolidates findings from various fields of study, the areas addressed are complex and multidisciplinary, and yet, little unified research exists on them. The study largely depends on the assumptions and theoretical propositions from studies conducted on new media platforms, making the findings highly contingent and may not be universally applicable to the metaverse due to its potent, unprecedented characteristics.

The concepts of media addiction and underlying psychological mechanisms in relation to metaverse are largely theoretical. Though such discussions have taken place in respect to social media platforms and internet usage (Bojic, 2017; Andreassen et al., 2016), empirical research on metaverse addiction has yet to be conducted. Therefore, the actual impact and effect size of metaverse on addictive behaviors need more empirical evidence.

The rare body of research on the metaverse makes it challenging to confirm or refute hypotheses. Though some tentative conclusions regarding addictive behaviors were established, these inferences are extrapolations from traditional media literature and assumptions from current VR/AR research.

The emerging stage of metaverse development precludes the possibility of long-term studies. The effects of widespread metaverse use over extended periods are unknown. The full range of effects, both positive and negative, cannot be accurately predicted given the current knowledge and existing research papers.

Future research directions should focus on empirical studies that can evaluate the effect and relationship between metaverse use and addictive behaviors. Such studies should also consider the differentiating factors and demographic variables, such as age, cultural background, and psychological predispositions.

Examining current 3D virtual environments such as Roblox and Decentraland could provide valuable insights into the addictive potential of the metaverse. Studies have shown that these platforms' immersive and interactive features can lead to excessive engagement, particularly among younger users, highlighting the importance of considering all forms of metaverse access in addiction research (Jose Miguel et al., 2024).

A complex interplay of recommender systems, content personalization, and social polarization in the metaverse needs further exploration. Understanding how to balance diverse content exposure and personalization to foster creativity and reduce addictions and echo chambers is an essential question for future studies.

While our study contributes to a growing body of research aimed at understanding the potential impacts and challenges of metaverse, the reviewed literature points to a promising and challenging future for interdisciplinary research on the metaverse.

## References

Abd-Elsayed, A., Hussain, N., & Stanley, G. (2021). Combining Virtual Reality and Behavioral Health to Promote Pain Resiliency: Analysis of a Novel BioPsychoSocial Modality for Solving Pain in the Workplace. *Pain and Therapy*, *10*(2), 1731–1740. https://doi.org/10.1007/s40122-021-00333-1

Aboujaoude, E., Kuss, D. J., Yao, M. Z., & Leung, L. W. (2021). Editorial: Online Psychology Beyond Addiction and Gaming: A Global Look at Mental Health and Internet-Related Technologies. *Frontiers in Psychology*, *12*. https://doi.org/10.3389/fpsyg.2021.815013

Ali, K., & Zain-ul-abdin, K. (2021). Post-truth propaganda: Heuristic processing of political fake news on Facebook during the 2016 U.S. presidential election. *Journal of Applied Communication Research*, *49*(1), 109–128. https://doi.org/10.1080/00909882.2020.1847311

Alontseva, D. V., & Chechurin, A. M. (2023). Education management in the context of the development of the digital educational environment: problems of the realization of the rights of minors to education. *Perspektivy Nauki i Obrazovania*, *61*(1), 672–689. https://doi.org/10.32744/pse.2023.1.40Andreassen, C. S., et al. (2016). The relationship between addictive use of social media and video games and symptoms of psychiatric disorders: A large-scale cross-sectional study. *Psychology of Addictive Behaviors*, *30*(2), 252-262.

Bailenson, J. N. (2021). Nonverbal overload: A theoretical argument for the causes of Zoom fatigue. *Technology, Mind, and Behavior*, *2*(1).

Billieux, J., Van Der Linden, M., Achab, S., Khazaal, Y., Paraskevopoulos, L., Zullino, D., & Thorens, G. (2013). Why do you play World of Warcraft? An in-depth exploration of self-reported motivations to play online and in-game behaviours in the virtual world of Azeroth. *Computers in Human Behavior*, *29*(1), 103–109. https://doi.org/10.1016/j.chb.2012.07.021

Biocca, Z., & Biocca, F. (2002). Building bridges across fields, universities, and countries: Successfully funding communication research through interdisciplinary collaboration. *Journal of Applied Communication Research*, *30*(4), 350–357. https://doi.org/10.1080/00909880216598

Bojic, L. (2022). Metaverse through the prism of power and addiction: What will happen when the virtual world becomes more attractive than reality? *European Journal of Futures Research*, *10*(1), 22. https://doi.org/10.1186/s40309-022-00208-4

Bojic, L. (2024). AI alignment: Assessing the global impact of recommender systems. *Futures*, *160*, 103383. https://doi.org/10.1016/j.futures.2024.103383Bojic, L. (2022). Metaverse through the prism of power and addiction: what will happen when the virtual world becomes more attractive than reality? *European Journal of Futures Research*, *10*(1).

Bojic, L., Agatonović, M., Guga, J. (2024). The Immersion in the Metaverse: Cognitive Load and Addiction. In: Geroimenko, V. (Ed) *Augmented and Virtual Reality in the Metaverse*. Springer Series on Cultural Computing. Springer, Cham. https://doi.org/10.1007/978-3-031-57746-8_11

Bojic, L., & Marie, J.-L. (2017). Addiction to Old versus New media. *Srpska politička misao*, *56*(2), 33-48. https://doi.org/10.22182/spm.5622017.2

Bojic, L., Dodevska, Z. Deldjoo, Y. & Nenad Pantelic, N. (2024b). Towards Recommender Systems LLMs Playground (RecSysLLMsP): Exploring Polarization and Engagement in Simulated Social Networks. *ROEGEN Workshop, 18th ACM Conference on Recommender Systems RECSYS*, 15-18 October 2024, Bari, Italy. https://roegen-recsys2024.github.io/papers/recsys2024-workshops_paper_173.pdf

Bolger, R. K. (2021). Finding Wholes in the Metaverse: Posthuman Mystics as Agents of Evolutionary Contextualization. *Religions*, *12*(9), 768.

Bratosin, I. A., Păvăloiu, I. B., Caculidis-Tudor, D., Luca, A. I., Goga, N., Podina, I. R., & Vasilăteanu, A. (2021). Virtual reality tools for pain management. *Revue Roumaine Des Sciences Techniques Serie Electrotechnique et Energetique*, *66*(1), 63–68. Retrieved from https://www.scopus.com/inward/record.uri?eid=2-s2.0-85108446370&partnerID=40&md5=87daf921e6bacfe94ed380d2e5c0521c

Bratosin, I.-A., Pavaloiu, I.-B., Vasilateanu, A., Gavajiuc, D., Dragoi, G., & Goga, N. (2019). *Pain Relief using Virtual Reality*. Presented at the Proceedings of the 11th International Conference on Electronics, Computers and Artificial Intelligence, ECAI 2019. https://doi.org/10.1109/ECAI46879.2019.9041996

Bratosin, I. A., Pavaloiu, I. B., Vasilateanu, A., Goga, N., Dragoi, G., & Gavajuc, D. (2020). *Virtual reality therapy for pain*. Presented at the eLearning and Software for Education Conference. https://doi.org/10.12753/2066-026X-20-193

Bridgstock, R., Cunningham, S., Twigg, D. (2018). Creating with mobile augmented reality: How can artists and designers move from screen-based to tangible making? *Convergence*, *24*(6), 633-650.

Caballero-Julia, D., Martín-Lucas, J., & Andrade-Silva, L. E. (2024). Unpacking the relationship between screen use and educational outcomes in childhood: A systematic literature review. *Computers and Education*, *215*. https://doi.org/10.1016/j.compedu.2024.105049

Cao, X., Gong, M., Yu, L., & Dai, B. (2020). Exploring the mechanism of social media addiction: An empirical study from WeChat users. *Internet Research*, *30*(4), 1305–1328. https://doi.org/10.1108/INTR-08-2019-0347

Carr, N. (2011). The Shallows: What the Internet Is Doing to Our Brains. W. W. Norton.

Casale, S., Musicò, A., & Schimmenti, A. (2021). Beyond internalizing and externalizing symptoms: The association between body disconnection and the symptoms of Internet gaming disorder. *Addictive Behaviors*, *123*. https://doi.org/10.1016/j.addbeh.2021.107043

Chagas, V. (2023). Far-right memespheres and platform affordances: The effects of environmental opacity on the spread of extremist memes on Twitter and WhatsApp. *Journal of Applied Communication Research*, *51*(6), 702–719. https://doi.org/10.1080/00909882.2023.2290897

Chen, K., Chen, J. V., & Ross, W. H. (2010). Antecedents of online game dependency: The implications of multimedia realism and uses and Gratifications theory. *Journal of Database Management*, *21*(2), 69–99. https://doi.org/10.4018/jdm.2010040104

Chen, M. (2023). The philosophy of the metaverse. *Ethics and Information Technology*, *25*(3), 41. https://doi.org/10.1007/s10676-023-09714-w

Cheng, C., & Ebrahimi, O. V. (2023). Gamification: a Novel Approach to Mental Health Promotion. *Current Psychiatry Reports*, *25*(11), 577–586. https://doi.org/10.1007/s11920-023-01453-5

Cheremoshkina, L. V., & Dupina, S. A. (2022). Mnemic Abilities of Students with Different Degrees of Immersion in Virtual Reality. *Psikhologicheskii Zhurnal*, *43*(3), 102–113. https://doi.org/10.31857/S020595920020500-6

Cinelli, M., De Francisci Morales, G., Galeazzi, A., Quattrociocchi, W., & Starnini, M. (2021). The echo chamber effect on social media. *Proceedings of the National Academy of Sciences*, *118*(9). https://doi.org/10.1073/pnas.2023301118

Cordero Jr., D. A. (2024). What a wonderful world? The downside of the metaverse phenomenon. *Journal of Public Health*, *46*(1), e206–e207. https://doi.org/10.1093/pubmed/fdad177

Dai, Y., Kim, J. W., & Jia, W. (2022). Health pandemic in the era of (Mis)information: Examining the utility of using victim narrative and social endorsement of user-generated content to reduce panic buying in the U.S. *Journal of Applied Communication Research*, *50*(5), 551–571. https://doi.org/10.1080/00909882.2022.2043557

de Clara, L. (2023). Transformative effect of virtual reality and the metaverse. In *AI and Emotions in Digital Society* (pp. 255–273). https://doi.org/10.4018/979-8-3693-0802-8.ch012Dhir, A., Yossatorn, Y., Kaur, P., & Chen, S. (2018). Online social media fatigue and psychological wellbeing—A study of compulsive use, fear of missing out, fatigue, anxiety and depression. *International Journal of Information Management*, *40*, 141-152.

Di Pietro, R., & Cresci, S. (2021). Metaverse: Security and Privacy Issues. The 3rd IEEE International Conference on Trust, Privacy and Security in Intelligent Systems and Applications (TPS'21).

Dominick, J. R. (2018). The dynamics of mass communication: Media in transition. McGraw-Hill.

Ehret, F., Francillette, Y., Girard, B., & Menelas, B.-A. J. (2021). *Action-Centered Exposure Therapy Using a Serious Game to Help Individuals with Alcohol Use Disorder*. Presented at the Lecture Notes in Computer Science (including subseries Lecture Notes in Artificial Intelligence and Lecture Notes in Bioinformatics). https://doi.org/10.1007/978-3-030-88272-3_13

Eiler, T. J., Grünewald, A., Machulska, A., Klucken, T., Jahn, K., Niehaves, B., … Brück, R. (2019). *A Preliminary Evaluation of Transferring the Approach Avoidance Task into Virtual Reality* (Vol. 1011). Presented at the Advances in Intelligent Systems and Computing. https://doi.org/10.1007/978-3-030-23762-2_14

Eiler, T. J., Grünewald, A., Wahl, M., & Brück, R. (2020). *AAT Meets Virtual Reality*. Presented at the Communications in Computer and Information Science. https://doi.org/10.1007/978-3-030-41590-7_7

Eiler, T. J., Schmücker, V., Haßler, B., Machulska, A., Grünewald, A., Klucken, T., … Brück, R. (2021). *Improving the Approach-Avoidance Task in Virtual Reality Through Presence and Virtual Risk Situations*. Presented at the Proceedings - 2021 4th IEEE International Conference on Artificial Intelligence and Virtual Reality, AIVR 2021. https://doi.org/10.1109/AIVR52153.2021.00041

Ellison, N. B., Steinfield, C., & Lampe, C. (2007). *The benefits of facebook "friends:" social capital and college students' use of online social network sites*. Journal of Computer-Mediated Communication, *12*(4), 1143–1168. https://doi.org/10.1111/j.1083-6101.2007.00367.x

Elsey, J. W. B., van Andel, K., Kater, R. B., Reints, I. M., & Spiering, M. (2019). The impact of virtual reality versus 2D pornography on sexual arousal and presence. *Computers in Human Behavior*, *97*, 35–43. https://doi.org/10.1016/j.chb.2019.02.031

Fleder, D. M. & Hosanagar, K. (2009). Blockbuster Culture's Next Rise or Fall: The Impact of Recommender Systems on Sales Diversity. *Management Science*, *55*(5), 697-712.

Flew, T. (2014). *New media: An introduction*. Oxford University Press.

Floridi, L. (2013). *The Ethics of Information*. Oxford University Press.

Garimella, K., Weber, I., & De Choudhury, M. (2017). Quote RTs on Twitter: Usage of the New Feature for Political Discourse. *8th ACM Conference on Web Science*, 200-204.

Garrett, B., Taverner, T., Gromala, D., Tao, G., Cordingley, E., & Sun, C. (2018). Virtual reality clinical research: Promises and challenges. *JMIR Serious Games*, *6*(4). https://doi.org/10.2196/10839

Goggin, G. (2012). *Global mobile media*. Routledge.

Gonçalves, J. (2018). Aggression in news comments: How context and article topic shape user-generated content. *Journal of Applied Communication Research*, *46*(5), 604–620. https://doi.org/10.1080/00909882.2018.1529419

Granic, I., Lobel, A., & Engels, R. C. M. E. (2014). The benefits of playing video games. *American Psychologist*, *69*(1), 66-78.

Greenfield, D. N. (2015). The addictive properties of Internet usage. In *Internet Addiction* (pp. 43-68). Springer.

Greenfield, S. (2014). Mind Change: How Digital Technologies Are Leaving Their Mark on Our Brains. Random House.

Hagins, J. (1996). The inconvenient public interest: Policy challenges in the age of information. *Journal of Applied Communication Research*, *24*(2), 83–92. https://doi.org/10.1080/00909889609365443

Han, D., Bergs, Y., & Moorhouse, N. (2022). Virtual reality consumer experience escapes: preparing for the metaverse. *Virtual Reality*, *26*(4), 1443-1458.

Hanshans, C., Maisch, B., Zauner, J., Faust, M. M. R., Broll, L. M., & Karch, S. (2021). Virtual Therapeutics-Requirements to deliver value with virtual reality and biofeedback applications for alcohol addiction therapy (Vol. 7). Presented at the Current Directions in Biomedical Engineering. https://doi.org/10.1515/cdbme-2021-2021

Hilmes, M. (2013). Network nations: A transnational history of British and American Broadcasting. Routledge.

Hoffmann, B. (2017). Phonoholism – a new behavioral addiction. *Trakia Journal of Science*, *15*(4), 315–319. https://doi.org/10.15547/tjs.2017.04.007

Hsieh, H.-F., & Shannon, S. E. (2005). Three approaches to qualitative content analysis. *Qualitative Health Research*, *15*(9), 1277−1288. https://doi.org/10.1177/1049732305276687

Jose Miguel A., M., Denise Rhian R., P., Leeann Kyle J., P., Derik Connery F., G., John Roche, R., & Ardrian D. M. (2024). Perspective of Senior High School Students on the Benefits and Risk of Playing Roblox. *American Journal of Open University Education*, *1*(1), 26–35. https://scientificbulletin.com/index.php/AJOUP/article/view/30

Kaplan, A. M., & Haenlein, M. (2020). Rulers of the world, unite! The challenges and opportunities of artificial intelligence. *Business Horizons*, *63*(1), 37-50.

Kim, S., & Kim, E. (2023). Emergence of the metaverse and psychiatric concerns in children and adolescents. *Journal of the Korean Academy of Child and Adolescent Psychiatry*, *34*(4), 215–221. https://doi.org/10.5765/jkacap.230047

Koban, A., Stevic, A., & Matthes, J. (2023). A tale of two concepts: Differential temporal predictions of habitual and compulsive social media use concerning connection overload and sleep quality. *Journal of Computer-Mediated Communication*, *28*(2), zmac040. https://doi.org/10.1093/jcmc/zmac040

Krishnan, A., & Zhou, X. (2019). Modeling the effect of health antecedents and social media engagement on healthy eating and quality of life. *Journal of Applied Communication Research*, *47*(4), 365–380. https://doi.org/10.1080/00909882.2019.1654124

Kumar, A., Shankar, A., Mehrotra, A., Yaqub, M. Z., & A. Alzeiby, E. A. (2024). Unveiling the dark and scary side of metaverse: An in-depth qualitative investigation. *Journal of Enterprise Information Management*. https://doi.org/10.1108/JEIM-04-2024-0195

Kuss, D., & Griffiths, M. (2017). Social Networking Sites and Addiction: Ten Lessons Learned. *International Journal of Environmental Research and Public Health*, *14*(3), 311.

Langener, S., Klaassen, R., VanDerNagel, J., & Heylen, D. (2022). Immersive Virtual Reality Avatars for Embodiment Illusions in People with Mild to Borderline Intellectual Disability: User-Centered Development and Feasibility Study. *JMIR Serious Games*, *10*(4). https://doi.org/10.2196/39966

Langener, S., Kolkmeier, J., VanDerNagel, J., Klaassen, R., van Manen, J., & Heylen, D. (2023). Development of an Alcohol Refusal Training in Immersive Virtual Reality for Patients with Mild to Borderline Intellectual Disability and Alcohol Use Disorder: Cocreation With Experts in Addiction Care. *JMIR Formative Research*, *7*. https://doi.org/10.2196/42523

Lee, L. -H. (2022). The Digital Big Bang in the Metaverse Era. *2022 IEEE International Symposium on Mixed and Augmented Reality Adjunct* (ISMAR-Adjunct), 55-55. doi: 10.1109/ISMAR-Adjunct57072.2022.00020

Limone, P., & Toto, G. A. (2021). Psychological and emotional effects of digital technology on children in covid-19 pandemic. *Brain Sciences*, *11*(9), 1126. https://doi.org/10.3390/brainsci11091126

Lister, M., Dovey, J., Giddings, S., Grant, I., Kelly, K. (2009). *New Media: A Critical Introduction*. Routledge.

Lombard, M., & Ditton, T. (2006). At the heart of it all: The concept of presence. *Journal of Computer-Mediated Communication*, *3*(2), 0–0. https://doi.org/10.1111/j.1083-6101.1997.tb00072.x

Machulska, A., Eiler, T. J., Grünewald, A., Brück, R., Jahn, K., Niehaves, B., … Klucken, T. (2020). Promoting smoking abstinence in smokers willing to quit smoking through virtual reality-approach bias retraining: A study protocol for a randomized controlled trial. *Trials*, *21*(1). https://doi.org/10.1186/s13063-020-4098-5

Maples-Keller, J. L., Bunnell, B. E., Kim, S.-J., & Rothbaum, B. O. (2017). The use of virtual reality technology in the treatment of anxiety and other psychiatric disorders. *Harvard Review of Psychiatry*, *25*(3), 103–113. https://doi.org/10.1097/HRP.0000000000000138

Matthes, J., Karsay, K., Schmuck, D., Stevic, A. (2020). 'Too much to handle': Impact of mobile social networking sites on information overload, depressive symptoms, and well-being. *Computers in Human Behavior*, *105*, 106217. https://doi.org/10.1016/j.chb.2019.106217

Mazza, M., Kammler-Sücker, K., Leménager, T., Kiefer, F., & Lenz, B. (2021). Virtual reality: a powerful technology to provide novel insight into treatment mechanisms of addiction. *Translational Psychiatry*, *11*(1). https://doi.org/10.1038/s41398-021-01739-3

McLuhan, M. (1964). Understanding Media: The Extensions of Man. McGraw-Hill.

Meng, Q. (2022). Application of VR Technology in Brand IP Image Design. *Mathematical Problems in Engineering*, *2022*. https://doi.org/10.1155/2022/2397056

Meyrowitz, J. (1985). *No Sense of Place: The Impact of Electronic Media on Social Behavior*. Oxford University Press.

Murray, J. (1999). *Hamlet on the Holodeck: The Future of Narrative in Cyberspace*. Cambridge, MA: MIT press. https://mitpress.mit.edu/9780262631877/hamlet-on-the-holodeck/

Mystakidis, S. (2022). *Metaverse*. Encyclopedia, *2*(1), 486-497.

Nadini, M., Alessandretti, L., Di Giacinto, F., Martino, M., Aiello, L. M., & Baronchelli, A. (2021). Mapping the NFT revolution: Market trends, trade networks, and visual features. *Scientific Reports*, *11*(1), 20902. https://doi.org/10.1038/s41598-021-00053-8

Naranjo-Zolotov, M., Turel, O., Oliveira, T., & Lascano, J. E. (2021). Drivers of online social media addiction in the context of public unrest: A sense of virtual community perspective. *Computers in Human Behavior*, *121*. https://doi.org/10.1016/j.chb.2021.106784

Ning, H. et al. (2021). *A Survey on Metaverse: The State-of-the-art, Technologies, Applications, and Challenges*. arXiv. https://doi.org/10.48550/arxiv.2111.09673

Olson, K. (2014). Essentials of Qualitative Interviewing. Routledge.

Ortiz De Gortari, A. B. (2016). The game transfer phenomena framework: Investigating altered perceptions, automatic mental processes and behaviors induced by virtual immersion. *Annual Review of CyberTherapy and Telemedicine*, *14*, 9–15. Retrieved from https://www.scopus.com/inward/record.uri?eid=2-s2.0-85019576435&partnerID=40&md5=a7974f8ebe9abb140f3caa96a2563cec

OSF (2024, May 10). Amplification of Addictive Features in the Metaverse. Retrieved from https://osf.io/mf3sa/?view_only=8db07d76bb244961a1f0e42ca6adb0e7

Pal, D., & Arpnikanondt, C. (2024). The sweet escape to metaverse: Exploring escapism, anxiety, and virtual place attachment. *Computers in Human Behavior*, *150*, 107998. https://doi.org/10.1016/j.chb.2023.107998

Pariser, E. (2012). *The Filter Bubble: How the New Personalized Web is Changing What We Read and How We Think*. Penguin Books.

Park, S., & Hwang, H. S. (2009). *Understanding online game addiction: Connection between presence and flow*. Presented at the Lecture Notes in Computer Science (including subseries Lecture Notes in Artificial Intelligence and Lecture Notes in Bioinformatics). https://doi.org/10.1007/978-3-642-02583-9_42

Park, S. M., & Kim, Y.-G. (2022). A Metaverse: Taxonomy, Components, Applications, and Open Challenges. *IEEE Access*, *10*, 4209-4251.

Peovic Vukovic, K. (2010). *Realism and virtuality: Carmageddon as contemporary simulacrum model*. Presented at the Lecture Notes in Computer Science (including subseries Lecture Notes in Artificial Intelligence and Lecture Notes in Bioinformatics). https://doi.org/10.1007/978-3-642-16638-9_39

Pira, F. (2024). The technological revolution, the metaverse and artificial intelligence. *Addiction & Social Media Communication*, *Vol 1*, 3-12 Pages. https://doi.org/10.13129/3035-1383/ASMC-4202

Pitts, M. W., Burnett, G., Skrypchuk, L., Wellings, T., Attridge, A., & Williams, M. (2012). Visual–haptic feedback interaction in automotive touchscreens. *Displays*, *33*(1), 7-16.

Przybylski, A. K., Murayama, K., DeHaan, C. R., & Gladwell, V. (2013). Motivational, emotional, and behavioral correlates of fear of missing out. *Computers in Human Behavior*, *29*(4), 1841-1848.

Ramesh, U., Harini, A., Gowri, D., Durga, K., Druvitha, P., & Kumar, K. S. (2022). Metaverse: Future of the Internet. *International Journal of Research Publication and Reviews*, *3*(2), 93-97.

Reiter, F., Heiss, R., Matthes, J. (2022). Explaining attitude-consistent exposure on social network sites: The role of ideology, political involvement, and network characteristics. *Social Science Computer Review*. Advance online publication. https://doi.org/10.1177/08944393211056224

Reyes, M. C. (2018). Measuring user experience on interactive fiction in cinematic virtual reality. In *Interactive Storytelling: 11th International Conference on Interactive Digital Storytelling, ICIDS 2018*, Dublin, Ireland, December 5–8, 2018, Proceedings 11 (pp. 295-307). Springer International Publishing. https://doi.org/10.1007/978-3-030-04028-4_33

Ricci, F., Rokach, L., & Shapira, B. (2010). *Introduction to Recommender Systems Handbook*. Springer eBooks, 1-35. doi: 10.1007/978-0-387-85820-3_1

Rick Faust, M. M., Hanshans, C., Broll, L. M., & Maisch, B. (2021). *Development of a virtual reality simulation for exposure based addiction therapy* (Vol. 7). Presented at the Current Directions in Biomedical Engineering. https://doi.org/10.1515/cdbme-2021-2022

Rieder, B. (2020). Engines of Order: *A Mechanology of Algorithmic Techniques*. Amsterdam University Press.

Riva, G., Wiederhold, B. K., Di Lernia, D., Chirico, A., Riva, E. F. M., Mantovani, F., … Gaggioli, A. (2019). Virtual reality meets artificial intelligence: The emergence of advanced digital therapeutics and digital biomarkers. *Annual Review of CyberTherapy and Telemedicine*, *17*, 3–7. Retrieved from https://www.scopus.com/inward/record.uri?eid=2-s2.0-85085609366&partnerID=40&md5=96fce33c0e8b3450c84334fd77801a83

Rózsa, S., Hargitai, R., Láng, A., Hartung, I., Martin, L., Tamás, I., … Kállai, J. (2020). Immersion into the mediated world: Adaptive and maladaptive consequences. *Mentalhigiene Es Pszichoszomatika*, *21*(4), 349–373. https://doi.org/10.1556/0406.21.2020.017

Ryan, M. L. (2014). Possible worlds. In Hühn, P., Meister, J. C., Pier, J., & Schmid, W. (Eds.) *Handbook of Narratology*, (pp. 726-742). Berlin: De Gruyter. https://doi.org/10.1515/9783110316469.726

Ryan, M. (2015). *Narrative as Virtual Reality 2: Revisiting Immersion and Interactivity in Literature and Electronic Media*. Baltimore: John Hopkins University Press. https://muse.jhu.edu/book/72246

Saker, M., & Frith, J. (2022). *Contiguous Identities*. First Monday.

Samala, A. D., Usmeldi, Taali, Ambiyar, Bojic, L., Indarta, Y., Tsoy, D., Denden, M., Tas, N., & Dewi, I. P. (2023). Metaverse Technologies in Education: A Systematic Literature Review Using PRISMA. International Journal of Emerging Technologies in Learning (iJET), 18(5), 231–252. https://doi.org/10.3991/ijet.v18i05.35501

Samala, A. D., Ricci, M., Angel Rueda, C. J., Bojic, L., Ranuharja, F., & Agustiarmi, W. (2024). Exploring campus through web-based immersive adventures using virtual reality photography: A low-cost virtual tour experience. *International Journal of Online and Biomedical Engineering (iJOE)*, *20*(01), 104–127. https://doi.org/10.3991/ijoe.v20i01.44339

Scherbaum, S., Haber, P., Morley, K., Underhill, D., & Moustafa, A. A. (2018). Biased and less sensitive: A gamified approach to delay discounting in heroin addiction. *Journal of Clinical and Experimental Neuropsychology*, *40*(2), 139–150. https://doi.org/10.1080/13803395.2017.1324022

Scopus (2024, May 8). *Welcome to Scopus Preview*. Scopus. https://www.scopus.com/

Scudder, J. N. (1999). Influence, beliefs, appraisals, and affect: A test of appraisal theory in a mediated context. *Journal of Applied Communication Research*, *27*(3), 196–216. https://doi.org/10.1080/00909889909365536

Selivanov, V. V. (2021). MENTAL STATES OF A PERSON IN A DIDACTIC VR ENVIRONMENT. *Experimental Psychology (Russia)*, *14*(1), 20–28. https://doi.org/10.17759/exppsy.2021000002

Shin, Y.-B., Kim, J.-J., Kim, H., Kim, S.-J., Eom, H., Jung, Y. H., & Kim, E. (2021). Managing game-related conflict with parents of young adults with internet gaming disorder: Development and feasibility study of a virtual reality app. *JMIR Serious Games*, *9*(1). https://doi.org/10.2196/22494

Siles González, I., & Meléndez Moran, A. (2021). “The most aggressive of algorithms”: User awareness of and attachment to TikTok’s content personalization. *International Communication Association (ICA) Mayo 27-31, 2021*. https://kerwa.ucr.ac.cr/handle/10669/83230

Simon, J., Etienne, A.-M., Bouchard, S., & Quertemont, E. (2020). Alcohol Craving in Heavy and Occasional Alcohol Drinkers After Cue Exposure in a Virtual Environment: The Role of the Sense of Presence. *Frontiers in Human Neuroscience*, *14*. https://doi.org/10.3389/fnhum.2020.00124

Siriaraya, P., Visch, V., Van Dooren, M. M. M., & Spijkerman, R. (2018). *Learnings and challenges in designing gamifications for mental healthcare: The case study of the readysetgoals application*. Presented at the 2018 10th International Conference on Virtual Worlds and Games for Serious Applications, VS-Games 2018 - Proceedings. https://doi.org/10.1109/VS-Games.2018.8493430Sun, Y., Wang, N., Shen, X.-L., & Zhang, J. Y. (2015). Location information disclosure in location-based social network services: Privacy calculus, benefit structure, and gender differences. *Computers in Human Behavior*, *52*, 278-292.

Stephenson, N. (1992, January 1). *Snow Crash*. Spectra. http://books.google.ie/books?id=B0xbAAAAMAAJ&q=0-553-08853-X&dq=0-553-08853-X&hl=&cd=6&source=gbs_api

Syakir, M. A., Maharani, S. A., Putra, Y. E., Handayani, P. W., & Asih, S. R. (2024). The Influence of Augmented Reality Face Filter Addiction on Online Social Anxiety: A Stimulus-Organism-Response Perspective. *Interdisciplinary Journal of Information, Knowledge, and Management*, *19*. https://doi.org/10.28945/5269Takac, C., Hvorecky, J., & Vojtek, P. (2018). Transfer collaborative filtering from multiple sources via consensus regularization. *International Conference on Intelligent Data Engineering and Automated Learning*, 559-568. Springer, Cham.

Tandoc, E. C., Ferrucci, P., & Duffy, M. (2015). Facebook use, envy, and depression among college students: Is facebooking depressing? *Computers in Human Behavior*, *43*, 139-146.

Tao, W., Hong, C., Tsai, W.-H. S., & Yook, B. (2021). Publics’ communication on controversial sociopolitical issues: Extending the situational theory of problem solving. *Journal of Applied Communication Research*, *49*(1), 44–65. https://doi.org/10.1080/00909882.2020.1849770

Tavinor, G. (2010). Virtual worlds and interactive fictions. In F. Lihoreau (Ed.), *Truth in Fiction* (pp. 223–244). Berlin: De Gruyter. https://doi.org/10.1515/9783110326796.223

Toronto, E. (2009). TIME OUT OF MIND: Dissociation in the Virtual World. *Psychoanalytic Psychology*, *26*(2), 117–133. https://doi.org/10.1037/a0015485Turel, O., & Serenko, A. (2012). The benefits and dangers of enjoyment with social networking websites. *European Journal of Information Systems*, *21*(5), 512-528.

Tracy, K., & Ashcraft, C. (2001). Crafting policies about controversial values: How wording disputes manage a group dilemma. *Journal of Applied Communication Research*, *29*(4), 297–316. https://doi.org/10.1080/00909880128115

Turillazzi, A., Taddeo, M., Floridi, L., & Casolari, F. (2023). The digital services act: An analysis of its ethical, legal, and social implications. *Law, Innovation and Technology*, *15*(1), 83–106. https://doi.org/10.1080/17579961.2023.2184136

Vanden Abeele, M. M., & Mohr, V. (2021). Media addictions as Apparatgeist: What discourse on TV and smartphone addiction reveals about society. *Convergence: The International Journal of Research into New Media Technologies*, *27*(6), 1536–1557. https://doi.org/10.1177/13548565211038539

van Gisbergen, M. S., Obermair, K., Iasonna, L., Doicaru, M., Rashnoodi, S. R., Campos, F., & Buwalda, V. J. A. (2024). REALISM IN RECOVRY: The Effect of Recorded (360°) and Computer Generated (CG) Environments and Humans in Virtual Reality Exposure Therapy (VRET). Presented at the Springer Proceedings in Business and Economics. https://doi.org/10.1007/978-3-031-50559-1_5

Veale, M., & Zuiderveen Borgesius, F. (2021). Demystifying the Draft EU Artificial Intelligence Act—Analysing the good, the bad, and the unclear elements of the proposed approach. *Computer Law Review International*, *22*(4), 97–112. https://doi.org/10.9785/cri-2021-220402

Vogel, E. R., Rose, J. P., Roberts, L. R., Eckles, K. (2014). Social comparison, social media, and self-esteem. *Psychology of Popular Media Culture*, 3(4), 206-222.

Zeng, R., Geng, L., Chen, H. (2017). *The Impacts of Recommender Systems on Sales Volume and Diversity*. arXiv preprint arXiv:1707.07642.

Zhai, X., Asmi, F., Zhou, R., Ahmad, I., Anwar, M. A., Saneinia, S., & Li, M. (2020). Investigating the Mediation and Moderation Effect of Students' Addiction to Virtual Reality Games: A Perspective of Structural Equation Modeling. *Discrete Dynamics in Nature and Society*, *2020*. https://doi.org/10.1155/2020/5714546

Zhang, X., Wu, Y., & Liu, S. (2019). Exploring short-form video application addiction: Socio-technical and attachment perspectives. *Telematics and Informatics*, *42*, 101243. https://doi.org/10.1016/j.tele.2019.101243

**Acknowledgements**

We would like to express our heartfelt gratitude to the Kooperation E-Medien Oesterreich (KEMÖ) Consortium for their support in getting this paper published.

This paper was realized with the support of the Ministry of Science, Technological Development and Innovation of the Republic of Serbia, according to the Agreement on the realization and financing of scientific research.

This research has been accomplished with the support and collaboration from the COST Action Network CA21129 - What are Opinions? Integrating Theory and Methods for Automatically Analyzing Opinionated Communication (OPINION) - https://www.opinion-network.eu/.

This paper has been supported by the TWON (project number 101095095), a research project funded by the European Union, under the Horizon Europe framework (HORIZON-CL2-2022-DEMOCRACY-01, topic 07). More details about the project can be found on its official website: https://www.twon-project.eu/.

**Availability of data and materials**

The data that support the findings of this study are available from the Open Science Framework (OSF). The dataset is openly accessible at https://osf.io/mf3sa/?view_only=8db07d76bb244961a1f0e42ca6adb0e7. The dataset is provided under a view-only link to ensure the integrity and traceability of the data. Researchers interested in exploring the dataset for further analysis or replication studies can access it directly via the aforementioned link.

**Competing interests**

The author declares that he has no competing interests.

**Funding**

This paper received no specific funding.

**Authors' contributions**

Ljubisa Bojic led the design and conceptualization of the study. His primary contribution involved an extensive literature review on the metaverse and its underlying technology. He orchestrated the exploration of potentially addictive features of new media platforms in the metaverse and contributed significantly to writing and editing the manuscript.

Jörg Matthes collaborated in study's design and contributed fundamentally to reviewing and interpreting the existing literature on the role of recommender systems. He provided meticulous insights on new media platforms' addictive features and their effects and played a key role in writing and revising the manuscript.

Agariadne Dwinggo Samala contributed to the analysis of papers and writing a final version of the manuscript.

Milan Cabarkapa brought specialized knowledge to the team, examining the potential impacts on creativity and social polarization from an engineering perspective. Primarily, he made significant contributions to the analysis and interpretation of existing literature, drafting the manuscript, and providing critical feedback.

Collectively, all authors critically evaluated the manuscript and approved the final version of the manuscript for submission.